\documentclass[10pt]{iopart}
\usepackage{iopams}  
\usepackage{graphicx}
\usepackage[breaklinks=true,colorlinks=true,linkcolor=blue,urlcolor=blue,citecolor=blue]{hyperref}
\usepackage{siunitx}
\usepackage[]{cite}
\usepackage{float}
\usepackage{subfig}

\begin{document}

\title[Contact resistance assessment and HF performance projection of BPFETs]{Contact resistance assessment and high-frequency performance projection of black phosphorus field-effect transistor technologies}\footnote{This is the Accepted Manuscript version of an article accepted for publication in \emph{Semiconductor Science and Technology}. IOP Publishing Ltd is not responsible for any errors or omissions in this version of the manuscript or any version derived from it. The Version of Record is available online at \href{https://iopscience.iop.org/article/10.1088/1361-6641/abbaed}{10.1088/1361-6641/abbaed}}

\author{Leslie M. Valdez-Sandoval$^{1}$, Eloy Ramirez-Garcia$^{1}$, David Jim\'enez$^{2}$, Anibal Pacheco-Sanchez$^{2}$}
\address{$^1$Instituto Polit\'ecnico Nacional, UPALM, Edif. Z-4 3er Piso, Cd. de M\'exico, 07738, M\'exico \\}
\address{$^2$Departament d'Enginyeria Electr\`{o}nica, Escola d'Enginyeria, Universitat Aut\`{o}noma de Barcelona, Bellaterra 08193, Spain \\}
\ead{lvaldezs1001@alumno.ipn.mx, ramirezg@ipn.mx, david.jimenez@uab.cat, anibaluriel.pacheco@uab.cat}

\begin{abstract}
\boldmath
In this work, an evaluation of the contact quality of black phosphorus (BP) field-effect transistors (FETs) from different technologies previously reported is performed by means of an efficient and reliable contact resistance extraction methodology based on individual device practical characteristics. A good agreement is achieved between the extracted values with the Y-function method used here and reference values obtained with other methods considering internal values as well as with more expensive methods involving fabricated test structures. The method enables a direct evaluation of different steps in the same technology and it embraces the temperature dependence of the contact characteristics. Channel phenomena have no impact on the extracted contact resistance values. High-frequency performance projections are obtained for fabricated devices based on the extracted contact resistance.
\end{abstract}

\vspace{2pc}
\noindent{\it Keywords}: Contact resistance, Schottky barrier, BPFET, AC performance

\section{Introduction}
\label{ch:intro}
Over the past decade, two dimensional (2D) semiconductors have emerged as promising candidates for future generations of nanoelectronic devices, due to their ultrathin bodies and high carrier mobility \cite{CaoW18,FioGia15}, that are considerably different from those in their bulk parental materials. More recently, black phosphorus (BP), with a direct bandgap of $\sim$ $\SI{0.3}{e\volt}$ in its bulk form and up to $\SI{2}{e\volt}$ for monolayers \cite{LiYu14}, has shown an excellent electronic efficiency for high performance transistors \cite{LiYu14,LiYu19}. Experiments have shown that, for certain device bias and temperature conditions, BP field-effect transistors (FETs) can exhibit an on/off ratio up to $\SI{10}{^6}$ and mobility up to $\SI{1000}{\centi\meter^2/\volt\cdot\second}$ \cite{LiYu14}. With its tunable band gap and high carrier mobility, BP is a suitable material to implement transistors for low-power high-frequency applications ~\cite{ZhuPar16,WanWan14,LiTia18}.

One of the major challenges in understanding and exploiting the intrinsic charge transport properties in emerging transistor technologies, such as BPFETs, arises from the contact resistance ($R_{\rm{C}}$) associated to interfaces between metal and low-dimensional channels such as 2D atomic layers. $R_{\rm{C}}$ is often associated to an energy- and material-dependent potential barrier induced by the interaction between the source and drain contacts and the two-dimensional channel material, as is the case for BP \cite{AshPen15}. In these Schottky-like FETs, it is important to understand the contact properties before extracting intrinsic properties of the channel such as a channel resistance $R_{\rm{ch}}$ describing the transport phenomena within the device channel. Hence, a reliable and $R_{\rm{ch}}$-independent characterization of $R_{\rm{C}}$ is required.

In general, the total contact resistance $R_{\rm{C}}$ embraces the contribution of the source contact resistance $R_{\rm{C,S}}$ and the drain contact resistance $R_{\rm{C,D}}$. In order to ease the study, these resistances are lumped here in a symmetrical dispossal such as $R_{\rm{C}}$ = $R_{\rm{C,S}}$ + $R_{\rm{C,D}}$. The total device resistance $R_{\rm{tot}}(= {V_{\rm{DS}}}/{I_{\rm{D}}})$ is the sum of channel and contact resistances, i.e., $R_{\rm{tot}}= R_{\rm{ch}}+R_{\rm{C}}$.

In this work, an $R_{\rm{C}}$-extraction methodology based on individual device characteristics is presented in Section \ref{ch:extract}. Contacts of fabricated BPFET technologies are characterized by extracting their corresponding $R_{\rm{C}}$ in Section \ref{ch:exp}. $R_{\rm{C}}$-enabled discussions regarding the impact of temperature and doping on potential barriers of metal-channel interfaces of some of the studied BPFETs are also included in Section \ref{ch:exp}. A high-frequency performance projection of fabricated BPFETs, enabled by $R_{\rm{C}}$ and other device parameters, is also presented. The final part of the work draws some conclusions. 

\section{Contact resistance extraction} \label{ch:extract}
In the literature, values for $R_{\rm{C}}$ of BPFETs have been obtained either by means of the fabrication of special test structures  \cite{LiYu19,HarMat15,YanCha17,XiaYin15} or by describing the device behaviour with adjusted models ~\cite{YarHar19,YarHar20,YarHar19_II}. The latter is a technology-specific approach which relies on the fitting parameters of an analytical or compact model (CM) and on the physical thoroughness of the model. The use of conventional extraction techniques in BPFETs, such as 4-point-probe (4PP) methods \cite{XiaYin15} and the transfer length method (TLM) \cite{LiYu19,HarMat15,YanCha17}, provides $R_{\rm{C}}$ values at the cost of additional process steps and whose reliability depends on a high-yield not reached yet by this emerging technology. 

The drift-diffusion-based $Y$-function \cite{Ghi88} has been used to extract values of $R_{\rm{C}}$ from individual device characteristics of different 2D ~\cite{ChaZhu14,ParSon18,PacFei20} emerging transistor technologies and without the need of additional test structures. $Y$-function-based methods rely on the relation between the drain current $I_{\rm{D}}$ in the linear regime and the square root of the transconductance $g_{\rm{m}}(=\partial I_{\rm{D}}/\partial V_{\rm{GS}})$ of a device such as $Y=I_{\rm{D}}/\sqrt{g_{\rm{m}}}$.

In this work, the electron drain current ($I_{\rm{D}}$) at the linear operation is considered as \cite{PacFei20,PacCla16}

\begin{equation} 
I_{\rm{D}} \approx \frac{\beta V_{\rm{GS,eff}}}{1+\theta V_{\rm{GS,eff}} } V_{\rm{DS}},
\label{eq:YFM_Id}
\end{equation}

\noindent where $V_{\rm{GS,eff}} = V_{\rm{GS}} - V_{\rm{th}} - V_{\rm{DS}}/2$ is the effective gate-to-source voltage with $V_{\rm{GS/DS}}$, the gate-to-source/drain-to-source voltage, $V_{\rm{th}}$ the threshold voltage, $\theta=\theta_0 + R_{\rm{C}}\beta$ is the extrinsic mobility degradation coefficient \cite{Ghi88,HaoCab85}, embracing the mobility degradation inside the channel due to vertical fields $\theta_0$,  $\beta=\mu_0 C_{\rm{ox}} w_{\rm{g}} / L_{\rm{g}}$ with $\mu_0$ as the low-field mobility, $C_{\rm{ox}}$ the oxide capacitance, and $w_{\rm{g}}$, and $L_{\rm{g}}$ the gate width and gate length, respectively. 

From Eq. (\ref{eq:YFM_Id}) and by using the $Y$-function and an auxiliar $X$-function ($X=1/\sqrt{g_{\rm{m}}}$), a bias-dependent contact resistance can be obtained as \cite{PacCla20}:

\footnotesize

\begin{equation}
	R_{\rm{C}} = \frac{V_{\rm{DS}}}{Y^2}  V_{\rm{GS,eff}}^2  \left[ \left( \frac{XY}{V_{\rm{GS,eff}}} - 1 \right) \left( \frac{1}{V_{\rm{GS,eff}}} \right) - \theta_0 \right].
    \label{eq:Rc_YFM_Vg}
\end{equation}

\normalsize
\noindent Notice that in contrast to other works \cite{ZhuPar16},\cite{ParSon18} where a  different $Y$-function based extraction method (YFM) has been used in BPFETs, $R_{\rm{C}}$ extracted here considers the effect of $\theta_0$, as well as a more complete model for $I_{\rm{D}}$, and hence, more accurate and complete information can be obtained by using Eq. (\ref{eq:Rc_YFM_Vg}) \cite{{PacCla16},{PacJim20}}. 

\section{Results and discussion} \label{ch:exp}
The development of BPFET technology has been demonstrated by different groups in the literature with fabricated proof-of-concept devices~\cite{LiYu19,ZhuPar16,WanWan14,LiTia18,HarMat15,YanCha17,HarNam16,XiaYin15,CheKua20}. In this section, the YFM discussed above has been applied in order to characterize the contacts of these transistors.

\subsection{$R_C$ characterization of fabricated devices} \label{ch:rc_cha}

The contacts of different BPFET technologies ~\cite{LiYu19,ZhuPar16,WanWan14,LiTia18,HarMat15, YanCha17} have been evaluated by extracting the corresponding $R_{\rm{C}}$ values with the YFM discussed above. Table \ref{tab:rc_exp} lists the device geometry of some of the studied technologies ~\cite{LiYu19,ZhuPar16,HarMat15,YanCha17,HarNam16}, as well as the reference contact resistivity $R_{\rm{C,ref}}\cdot w_{\rm{g}}$, with $w_{\rm{g}}$ as the device gate width, obtained with other methods as reported in the corresponding reference. Notice that a different technology implies different device footprints, e.g. gate length $L_{\rm{g}}$, architectures and fabricated processes, and hence, a systematic scaling study is not feasible in this work despite the universality of the method presented here. However, doping- and temperature-dependent $R_{\rm{C}}$ studies are presented (cf. Figs. \ref{fig:R_YFM_r9} and \ref{fig:RC_YFM}).

\Table{Device dimensions and reference contact resistivity of fabricated BPFETs.}
\br
[ref.]&$w_{\rm{g}}$&${L_{\rm{g}}}$&${R_{\rm{C,ref}}}\cdot w_{\rm{g}}$&extraction\\
&(\si{\micro\meter})&(\si{\nano\meter})&(\si{\kilo\ohm \cdot \micro\meter})&method\\
\mr
\mr
\cite{LiYu19} & \SI{2.3}  & \SI{100}{}  & \SI{0.7}{} &  TLM \\ &&&($V_{\rm{GS}}=\SI{-4}{\volt}$,\\ &&&$V_{\rm{DS}}$ not reported)  \\ \mr 
\cite{ZhuPar16} & \SI{11}{} & \SI{250}{} & \SI{4.5}{} - \SI{6.7}{} &simplified\\ &&&(bias not reported) & YFM \\ \mr
\cite{HarMat15}  & --  & \SI{170}{} & \SI{2.28}{} & TLM\\ &&&($V_{\rm{GS}}=-\SI{1.5}{\volt}$,\\ &&&$V_{\rm{DS}}$ not reported)  \\ \mr
\cite{YanCha17} & \SI{10}{}  & \SI{200}{} & \SI{0.58}{} & TLM \\ &&&($V_{\rm{GS}}=\SI{0}{\volt}$,\\ &&&$V_{\rm{DS}}=-\SI{1}{\volt}$)  \\ \mr
\cite{HarNam16} & \SI{3.16}{}  & \SI{300}{} & \SI{1.4}{}  & CM in \cite{YarHar19_II} \\ &&&(at linear region) \\ \mr
\cite{XiaYin15} & \SI{3}{}  & \SI{}{}$-$ & \SI{101.02}{}  & 4PP \\ &&&($V_{\rm{GS}}=-\SI{54.4}{\volt}$,\\ &&&$V_{\rm{DS}}$ not reported) \\
\br
\end{tabular}
\label{tab:rc_exp}
\end{indented}
\end{table}

\begin{figure}[!htb]
\centering
\includegraphics[scale=0.3]{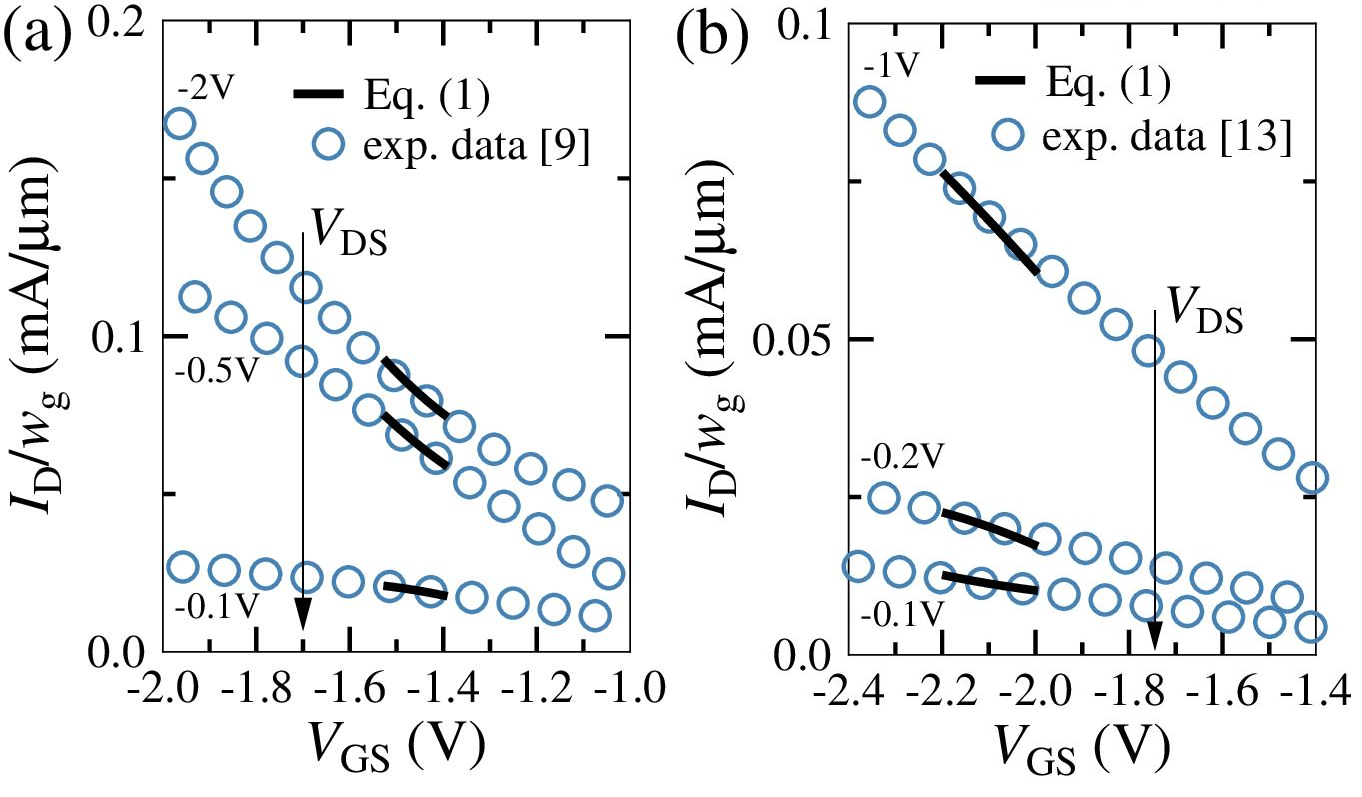}
	\caption{Transfer characteristics of the BPFETs: \textbf{(a)} for a \SI{170}{\nano\meter}-long device \cite{HarMat15}, \textbf{(b)} for a \SI{300}{\nano\meter}-long device \cite{YarHar19_II}.}
\label{fig:C_trans_vs_YFM}
\end{figure}

The experimental transfer characteristics of some of the devices under study (DUT) \cite{HarMat15,YarHar19_II} are shown in Fig. \ref{fig:C_trans_vs_YFM}. The drain current obtained by Eq. (\ref{eq:YFM_Id}) with the extracted parameters is also shown within the bias range in which YFM has been applied in each device. The good match between experimental data and Eq. (\ref{eq:YFM_Id}) indicates the validity of the extracted parameters. Similar results have been obtained for the other devices studied in this work. According to the authors knowledge this verification step has not been previously reported for BPFETs.

\begin{figure}[!htb]
	\centering
	\includegraphics[scale=0.3]{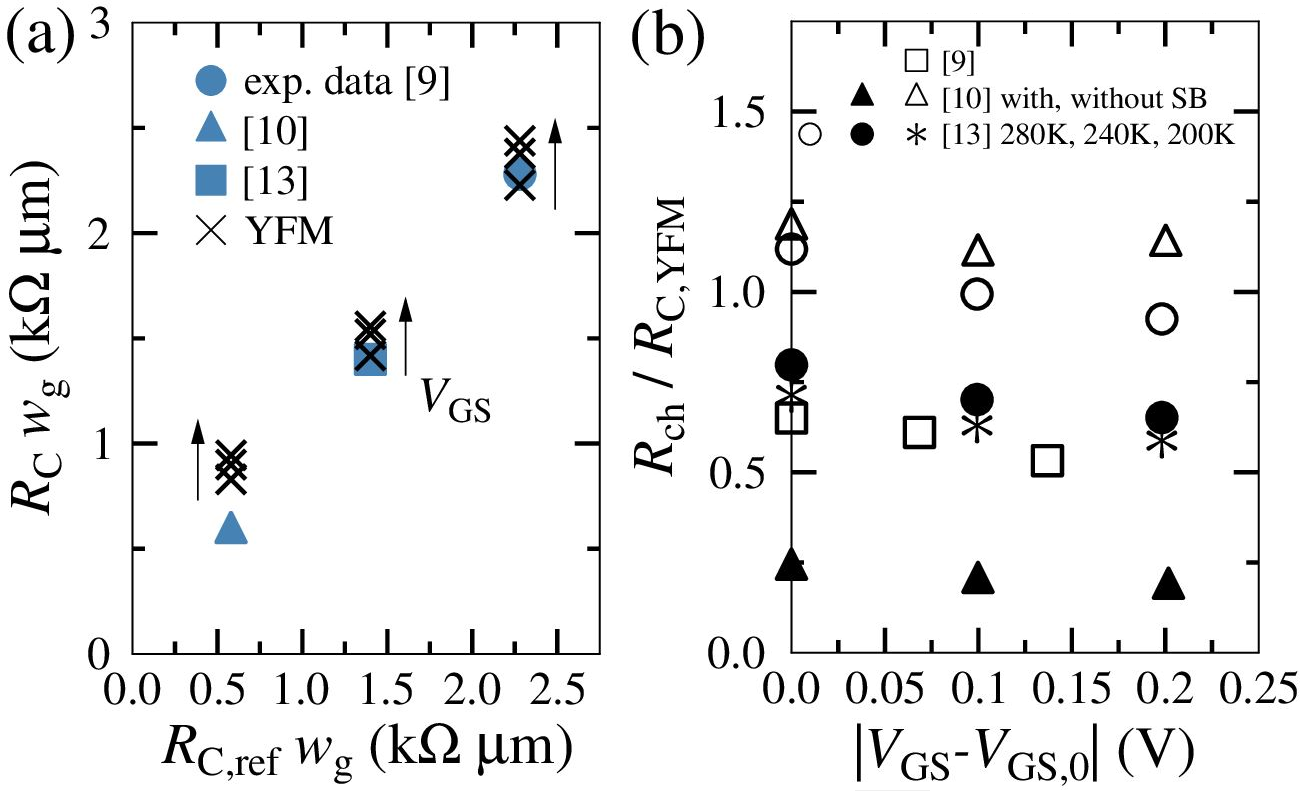}
	\caption{\textbf{(a)} Contact resistivity of different BPFET technologies: reference values have been reported from TLM measurement for a \SI{170}{\nano\meter}-long device \cite{HarMat15} and for a \SI{200}{\nano\meter}-long device \cite{YanCha17}, and from an adjustment of a CM for a \SI{300}{\nano\meter}-long device \cite{YarHar19_II}. All crosses correspond to $R_{\rm{C,YFM}}$ values extracted at $V_{\rm{DS}}=\SI{-0.1}{\volt}$ and different $V_{\rm{GS}}$: for \cite{HarMat15} the bias range is $\SI{-1.5}{\volt}\leq V_{\rm{GS}}\leq \SI{-1.39}{\volt}$, for \cite{YanCha17} is $\SI{-0.2}{\volt}\leq V_{\rm{GS}}\leq \SI{0.1}{\volt}$ and for \cite{YarHar19_II} is $\SI{-2.5}{\volt}\leq V_{\rm{GS}}\leq \SI{-2.3}{\volt}$. \textbf{(b)} Ratio between channel resistance ($R_{\rm{ch}}$) and $R_{\rm{C,YFM}}$ \cite{HarMat15,YanCha17,YarHar19_II}, all reported $R_{\rm{ch}}$/$R_{\rm{C,YFM}}$ ratios correspond to $V_{\rm{DS}}=\SI{-0.1}{\volt}$. Extracted values of \cite{YarHar19_II} are also reported at different temperatures.}
	\label{fig:Rch_RC}
\end{figure}

\begin{figure}[!htb]
\begin{center}
	\includegraphics[scale=0.4]{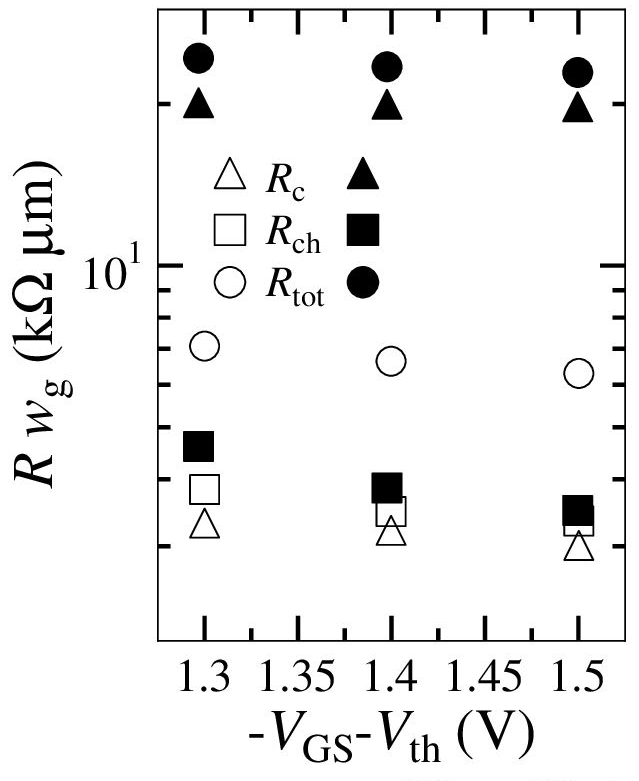}
	\caption{Contact resistivity, channel resistivity, and device total resistivity of the \SI{200}{\nano\meter}-long transistor, $R_{\rm{tot}}$ has been extracted from transfer curves in \cite{YanCha17} at $V_{\rm{DS}}=\SI{-0.1}{\volt}$. Unfilled markers correspond to a \SI{7}{\nano\meter} thick BP-PMOSFET without boron nitride (BN) tunneling barrier at source/drain. Filled markers correspond to the same device but with bilayer BN tunneling barriers at source and drain.}
\label{fig:R_YFM_r9}
\end{center}
\end{figure}

\begin{figure}[!htb]
\centering
\includegraphics[scale=0.31]{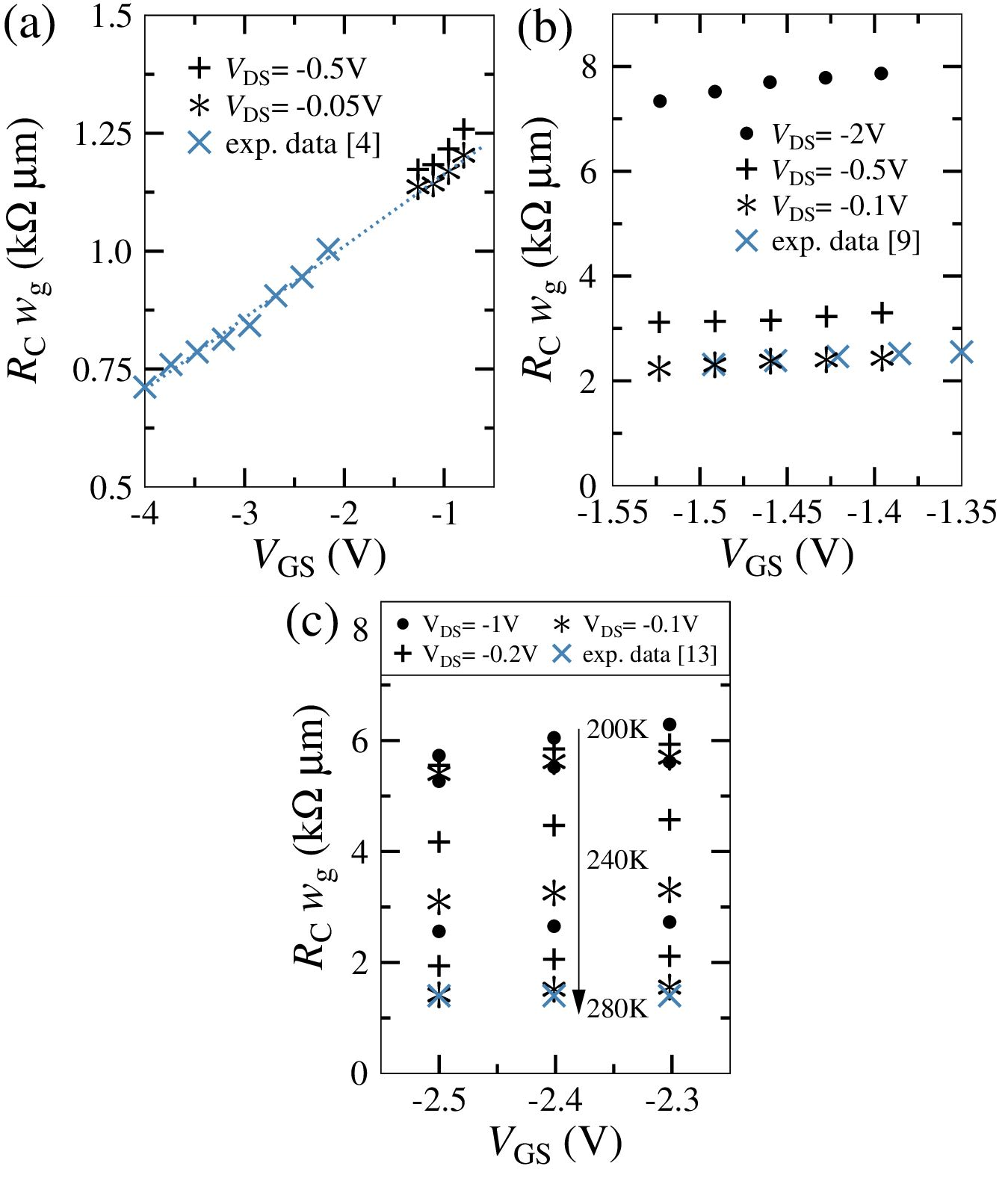}
	\caption{Contact resistivity of the BP device calculated by Eq. (\ref{eq:Rc_YFM_Vg}) from transfer characteristics. \textbf{(a)} for a \SI{100}{\nano\meter}-long device \cite{LiYu19}, dotted line represents a linear extrapolation, \textbf{(b)} for a \SI{170}{\nano\meter}-long device \cite{HarMat15}, \textbf{(c)} for a \SI{300}{\nano\meter}-long device \cite{YarHar19_II} at different temperatures.}
\label{fig:RC_YFM}
\end{figure}

Fig. \ref{fig:Rch_RC} (a) highlights the $R_{\rm{C}}$ extracted values of the different DUTs \cite{HarMat15,YanCha17,YarHar19_II} at different $V_{\rm{GS}}$. In contrast to the reference values, $R_{\rm{C,YFM}}$ shows a bias-dependence, due to the contact characteristics, i.e., modulation of potential contact barriers by an electric field. In addition, $R_{\rm{C,YFM}}$ extracted values are close to the reference values extracted using the TLM-method and CM-method (filled markers in Fig. \ref{fig:Rch_RC} (a)) \cite{HarMat15,YanCha17,YarHar19_II}. The slight differences between extracted and reference values can be due to an extraction under different bias conditions (see Table I), e.g., the bias point or range at which reference $R_{\rm{C}}$ for the \SI{300}{\nano\meter}-long device \cite{YarHar19_II} is valid has not been reported.

The ratio between channel resistance and extracted contact resistance indicates the impact of $R_{\rm{ch}}$ and $R_{\rm{C}}$ on the device performance. Fig. \ref{fig:Rch_RC} (b) shows this ratio, for the DUTs \cite{HarMat15,YanCha17,YarHar19_II} over the bias range in which $R_{\rm{C}}$ has been extracted. $V_{\rm{GS,0}}$ is the $V_{\rm{GS}}$ closest to the threshold voltage. If the ratio $R_{\rm{ch}}$/$R_{\rm{C}}$ is close to 1, it means that both, $R_{\rm{ch}}$ and $R_{\rm{C}}$ contribute similarly to the device total resistance, which is the case for the DUTs included in Table \ref{tab:rc_exp}, \cite{HarMat15,YanCha17,YarHar19_II}. The impact of the contact or channel properties on the device performance can be quantified independently. This is embraced by the extraction method as demonstrated by analyzing the ratio of $R_{\rm{ch}}$ to the extracted $R_{\rm{C}}$ of the \SI{200}{\nano\meter} device \cite{YanCha17} with and without boron nitride (BN)-induced potential barriers at the source and drain contacts. As shown also in Fig. \ref{fig:Rch_RC} (b), $R_{\rm{C}}$ dominates the performance of the \SI{200}{\nano\meter}-long device \cite{YanCha17} with BN barriers, because they add an additional resistance to the metal-channel interface while the channel properties remain the same. Furthermore, Figs. \ref{fig:Rch_RC} (a) and (b) show that YFM is as reliable as other methods, however, in contrast to the latter, YFM allows the evaluation of different technologies from the $I-V$ characteristics and at different temperatures of individual devices without the need of test stuctures or an adjustment of a complete set of model parameters.

Fig. \ref{fig:R_YFM_r9} shows the contact ${R_{\rm{C}}}\cdot w_{\rm{g}}$, channel ${R_{\rm{ch}}}\cdot w_{\rm{g}}$ and total resistivity ${R_{\rm{tot}}}\cdot w_{\rm{g}}$, of the  \SI{200}{\nano\meter}-long devices with and without additional BN barriers at the contacts \cite{YanCha17}. Notice that $R_{\rm{C}}$ has been extracted  for each device at similar transistor operation regions with respect to $V_{\rm{th}}$ towards a fair comparison in terms of transport conditions. For both cases, $R_{\rm{tot}}$ has been obtained using $R_{\rm{tot}}= {V_{\rm{DS}}}/{I_{\rm{D}}}$, $R_{\rm{C}}$ by the YFM method, Eq. (\ref{eq:Rc_YFM_Vg})  and ${R_{\rm{ch}}}=R_{\rm{tot}}-R_{\rm{C}}$. $R_{\rm{C}}$ is higher for the device with additional barrier compared to the barrier-free device. Therefore $R_{\rm{tot}}$ also increases, which will cause an $I_{\rm{D}}$ decrease. The $R_{\rm{ch}}$  results are almost equal in both devices. This implies that (i) transport phenomena within the channel material have minimum impact on the device performance, in contrast to the phenomena associated with the characteristics of contacts and (ii) the extraction method used here is totally independent of channel phenomena.

Figs. \ref{fig:RC_YFM} (a)-(b) show the extracted $R_{\rm{C,YFM}}$ with reference values reported over a bias range of a \SI{100}{\nano\meter}-long device and a \SI{170}{\nano\meter}-long device. The reported values ($R_{\rm{C,TLM}}$) in Fig. \ref{fig:RC_YFM} (a), are evaluated at a $V_{\rm{GS}}$-bias range lower than the $V_{\rm{GS}}$-bias range in this paper, however, by extrapolating the incremental trend of $R_{\rm{C,TLM}}$ the values of $R_{\rm{C,YFM}}$ are obtained. The comparison between $R_{\rm{C,TLM}}$ measured in \cite{HarMat15} and $R_{\rm{C}}$ extracted by the YFM method at several $V_{\rm{GS}}$ is shown in Fig. \ref{fig:RC_YFM} (b), this comparison highlights that the difference between $R_{\rm{C,TLM}}$ and $R_{\rm{C,YFM}}$ is minimal within the similar bias range in which both methods have been applied. Fig. \ref{fig:RC_YFM} (c), shows the variation of $R_{\rm{C}}$ for a BP-device at different temperatures, and it can be concluded that the $R_{\rm{C}}$ decreases as the temperature increases, which coincides with an analytical model of a temperature-dependent contact resistance model shown elsewhere \cite{YarHar20}. Furthermore, the $V_{\rm{DS}}$-dependence of $R_{\rm{C}}$ reveals the sensitivity of phenomena within the metal-BP interface to lateral electric fields. The contribution at source and drain interfaces of these fields can be either symmetrical or unbalanced, however, this is out of the scope of this study.

The method has been proven to extract the contact resistivity, also for BPFETs with more challenging channel configurations, since $R_{\rm{ch}}$ has no impact on the $R_{\rm{C}}$ extraction in YFM, such as the device in \cite{XiaYin15} where the YFM-extracted contact resistivity is of \SI{107}{\kilo\ohm \cdot \micro\meter} which is $\approx$ 6$\%$ close to the reported value of \SI{101}{\kilo\ohm \cdot \micro\meter} (see Fig. 2(c) in \cite{XiaYin15}) for $V_{\rm{GS}}$ in the range of \SI{-53.1}{\volt} to \SI{-55.6}{\volt}.
	
\subsection{$R_C$-based high-frequency performance projection} \label{ch:A}

\begin{figure}[!htb]
	\begin{center}
		\includegraphics[scale=0.31]{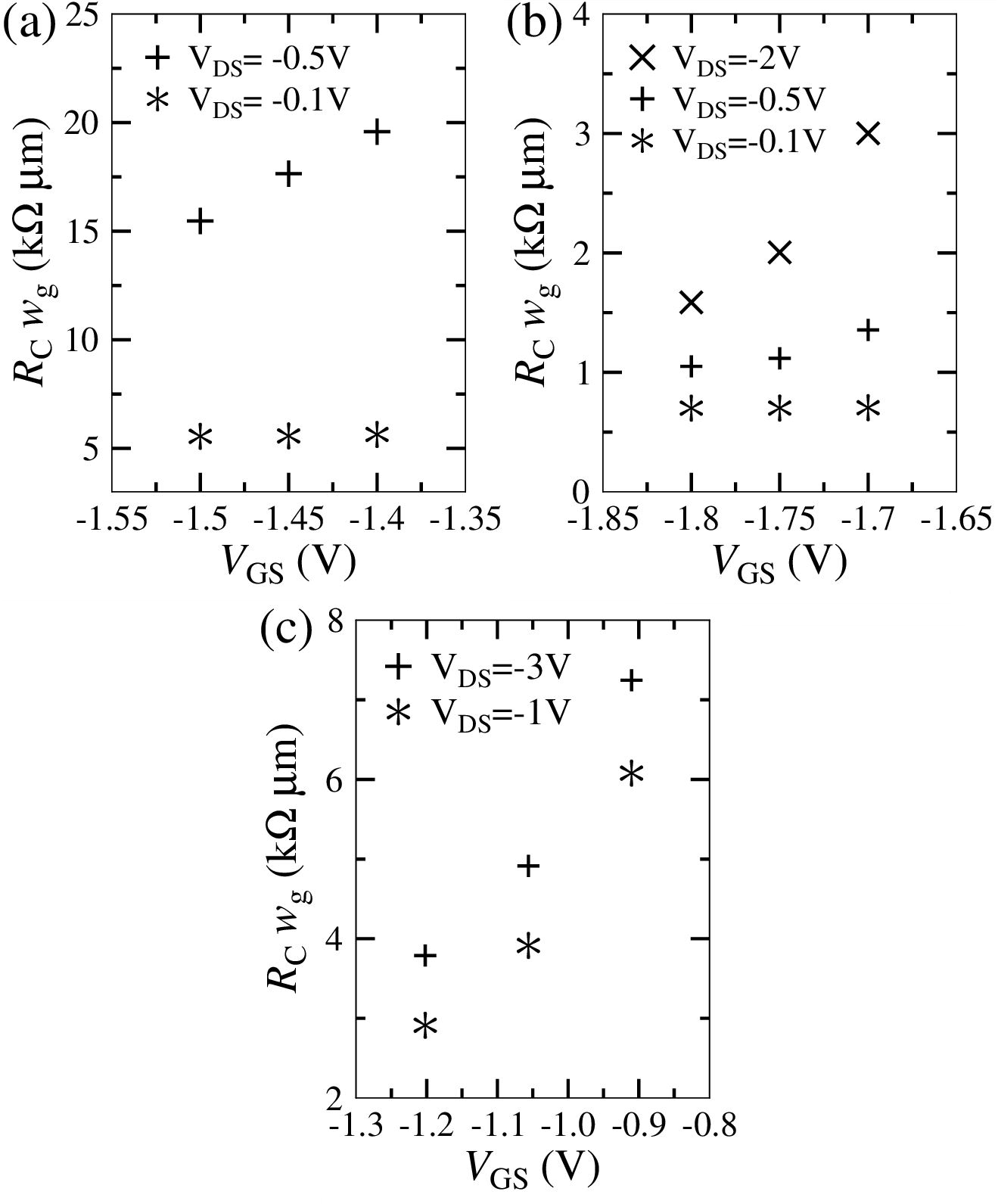}
		\caption{Contact resistivity of high frequency BP devices from different technologies: \textbf{(a)} a \SI{250}{\nano\meter}-long device \cite{ZhuPar16}, \textbf{(b)} a \SI{300}{\nano\meter}-long device \cite{WanWan14}, and \textbf{(c)} a \SI{400}{\nano\meter}-long device \cite{LiTia18}.}
		\label{fig:RC_YFM_HF}
	\end{center}
\end{figure}

The $R_{\rm{C}}$ extracted for high-frequency BP-FETs ~\cite{ZhuPar16,WanWan14,LiTia18} in Figs. \ref{fig:RC_YFM_HF} (a)-(c) show that $R_{\rm{C}}$ decreases as $V_{\rm{DS}}$ increases, as well as a larger variation with $V_{\rm{GS}}$. The latter embraces the potential barrier change due to vertical fields. The $R_{\rm{C}}$ values extracted from \cite{ZhuPar16} are shown in Fig. \ref{fig:RC_YFM_HF} (a), it can be seen that $R_{\rm{C}}$ extracted ($\approx$\SI{5.6}{\kilo\ohm \cdot \micro\meter}) at $V_{\rm{DS}} = \SI{-0.1}{\volt}$ is between the values reported in \cite{ZhuPar16} (see Table \ref{tab:rc_exp}), therefore, this is indicative that the YFM method obtains more accurate data with fewer simplifications, compared to the simplified YFM method used in \cite{ZhuPar16}.
%

The high-frequency performance of bias-dependent BPFETs can be described by an equivalent small-signal circuit model shown elsewhere \cite{PacFei20,YinAlM17}. For a symmetrical disposal of source contact resistance and drain contact resistance, i.e., $R_{\rm{C}}/\SI{2}=R_{\rm{s}}=R_{\rm{d}}$, the extrinsic cutoff frequency $f_{\rm{T,e}}$ and the extrinsic maximum oscillation frequency $f_{\rm{MAX,e}}$ are given by \cite{YinAlM17}

\normalsize
\begin{equation}
f_{\rm{T,e}} \approx \frac{g_{\rm{m,i}}} {2\pi \{ C_{\rm{gg,t}} [1+g_{\rm{d,i}} R_{\rm{C}}] + C_{\rm{gd,t}} (g_{\rm{m,i}} R_{\rm{C}}) \} },
\label{eq:fT}
\end{equation}

\begin{equation}
f_{\rm{MAX,e}} \approx \frac{g_{\rm{m,i}}} {4\pi \sqrt{\Psi_1+\Psi_2+\Psi_3}},
\label{eq:fMAX}
\end{equation}

\noindent where the total gate-to-source capacitance ($C_{\rm{gg,t}}$) and total gate-to-drain capacitance ($C_{\rm{gd,t}}$) have been obtained by a simple practical approach towards the evaluation of the impact of $R_{\rm{C}}$ over the HF performance, i.e., by using Eqs. (4) and (5) in \cite{YinAlM17}, the experimental intrinsic $f_{\rm{T}}$ and $f_{\rm{MAX}}$ reported in the corresponding references ~\cite{ZhuPar16,WanWan14,LiTia18,CheKua20} (see Table \ref{tab:FoMs}) and by assuming bias-independent capacitances within the bias range of interest for this study.

\begin{table}
	\centering
	\caption{FoMs of different technologies of BPFETs.}
	\begin{tabular}{c c c c}
		\br
		[ref.]&$f_{\rm{T,e}}$&${f_{\rm{MAX,e}}}$&${R_{\rm{C,ref}}}\cdot w_{\rm{g}}$\\
		&(\si{\giga\hertz})&(\si{\giga\hertz})&(\si{\kilo\ohm \cdot \micro\meter})\\ \mr
		\mr
		\cite{ZhuPar16} & \SI{6}{} & \SI{10.72}{} & \SI{4.5}{} - \SI{6.7}{}\\ &&&(bias not reported) \\ \mr
		\cite{WanWan14}  & \SI{8}{}  & \SI{12}{} & \SI{}{}- \\ \mr
		\cite{LiTia18} & \SI{2}{}  & \SI{17}{} & \SI{}{}-\\ \mr
		\cite{CheKua20} & \SI{37}{}  & \SI{22}{} & \SI{6}{}\\ &&&($V_{\rm{GS}}=-\SI{0.5}{\volt}$,\\ &&&$V_{\rm{DS}}=-\SI{1}{\volt}$) \\ \br
	\end{tabular}
	\label{tab:FoMs}
\end{table}

$R{\rm{g}}$ values have been obtained by using Eq. (6) in \cite{YinAlM17} considering the corresponding device geometry. Similarly, $\Psi_{\rm{1}}$, $\Psi_{\rm{2}}$, $\Psi_{\rm{3}}$ have been calculated using Eq. (9) in \cite{YinAlM17}. Table \ref{tab:Cg_param} shows the bias-independent calculated values for each device in this study. For the intrinsic transconductance $g_{\rm{m,i}}$(=$\partial$$I_{\rm{D}}$/$\partial$$V_{\rm{GS,i}}$), obtained from transfer characteristics, and intrinsic output conductance $g_{\rm{d,i}}$(=$\partial$$I_{\rm{D}}$/$\partial$$V_{\rm{DS,i}}$),  obtained from output characteristics, the intrinsic gate-to-source voltage $V_{\rm{GS,i}}$$\approx$$V_{\rm{GS}}$-$I_{\rm{D}}$$R_{\rm{C}}$/\SI{2}{} and intrinsic drain-to-source voltage $V_{\rm{DS,i}}$$\approx$$V_{\rm{DS}}$-$I_{\rm{D}}$$R_{\rm{C}}$, have been obtained by considering the extracted $R_{\rm{C,YFM}}$ for each device (cf. Fig. \ref{fig:RC_YFM_HF}).

\begin{table}
	\centering
	\caption{Capacitances and gate resistance of different technologies of BPFETs.}
	\begin{tabular}{c c c c}
		\br
		[ref.]&$C_{\rm{gg,t}}$&${C_{\rm{gd,t}}}$&${R_{\rm{g}}}$\\
		&(\si{\atto\farad})&(\si{\atto\farad})&(\si{\ohm})\\ \mr
		\mr
		\cite{ZhuPar16} & \SI{0.28}{} & \SI{0.13}{} & \SI{14.66}{}\\ \mr
		\cite{WanWan14}  & \SI{168.19}{}  & \SI{83.78}{} & \SI{12.22}{}\\ \mr
		\cite{LiTia18} & \SI{53.8}{}  & \SI{26.5}{} & \SI{10}{}\\ \br
	\end{tabular}
	\label{tab:Cg_param}
\end{table}

\begin{figure}[!htb]
\begin{center}
\includegraphics[scale=0.31]{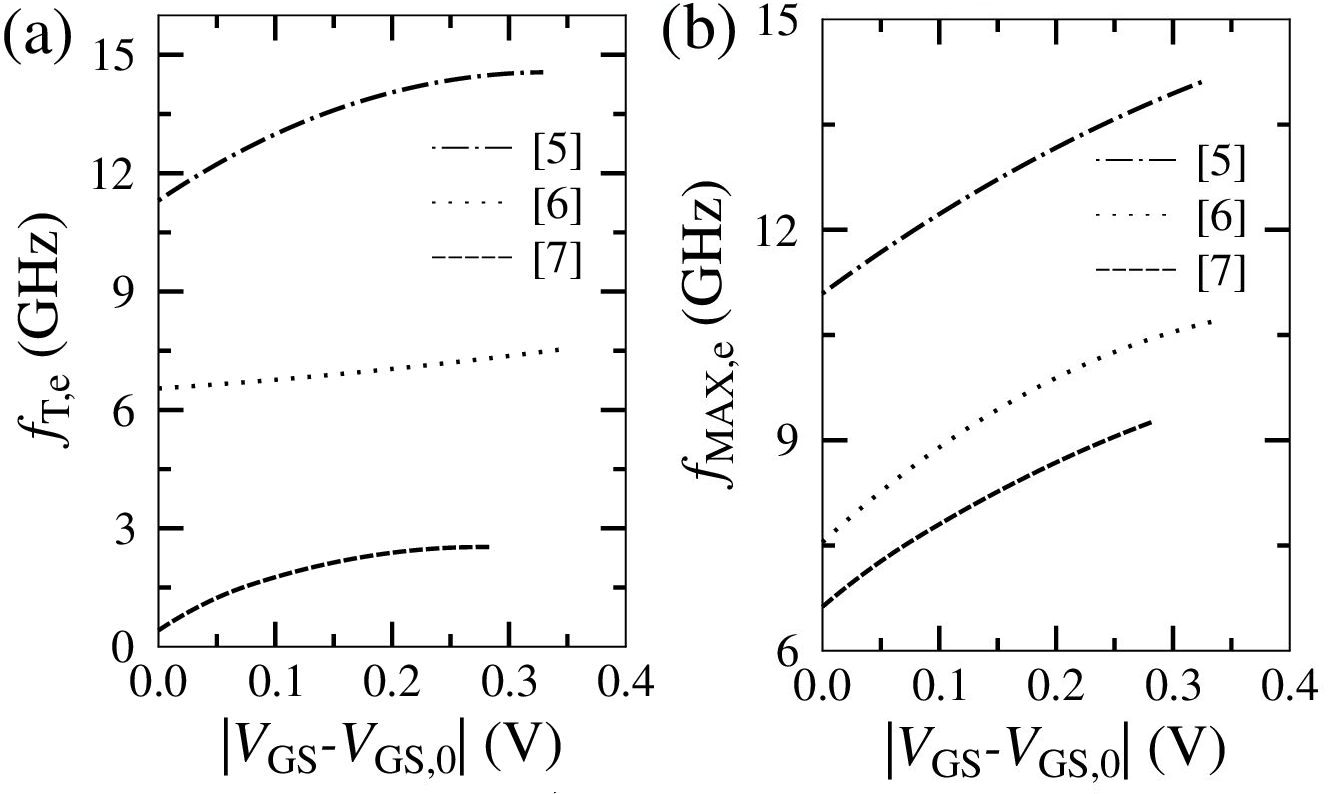}
	\caption{\textbf{(a)} Extrinsic cutoff frequency and \textbf{(b)} extrinsic maximum oscillation frequency within the bias range where $R_{\rm{C}}$ has been extracted: \SI{250}{\nano\meter}-long device \cite{ZhuPar16} ($V_{\rm{DS}}=\SI{-0.1}{\volt}$, $V_{\rm{GS,0}}=\SI{-1.4}{\volt}$), \SI{300}{\nano\meter}-long device \cite{WanWan14} ($V_{\rm{DS}}=\SI{-0.1}{\volt}$, $V_{\rm{GS,0}}=\SI{-1.7}{\volt}$) and \SI{400}{\nano\meter}-long device \cite{LiTia18} ($V_{\rm{DS}}=\SI{-1}{\volt}$, $V_{\rm{GS,0}}=\SI{-0.9}{\volt}$).}
\label{fig:fT_fMAX_HF}
\end{center}
\end{figure}

The high-frequency Figures of Merit (FoM) expressed by Eqs. (\ref{eq:fT}) and (\ref{eq:fMAX}) have been obtained for BPFET technologies for high-frequency applications: a \SI{250}{\nano\meter}-long device \cite{ZhuPar16}, a \SI{300}{\nano\meter}-long device \cite{WanWan14} and a \SI{400}{\nano\meter}-long device \cite{LiTia18}. FoMs are reported in Fig. \ref{fig:fT_fMAX_HF}. The FoMs in \cite{CheKua20} are not analyzed in this study because the YFM method is valid for a three-terminal device (gate, source and drain), while in \cite{CheKua20} two extra terminals are considered (five contacts in total) to induce electrostatic doping both in the source and drain, however, we report these FoMs because the device described in \cite{CheKua20} achieved the best $f_{\rm{T}}$/$f_{\rm{MAX}}$ performances for a BPFET technology.

Notice that ($f_{\rm{T,e}}$, $f_{\rm{MAX,e}}$), inferred from $R_{\rm{C,YFM}}$, roughly approximate the reported values of (6, 10.72) GHz \cite{ZhuPar16}, (8, 12) GHz \cite{WanWan14} and (2, 17) GHz \cite{LiTia18}. The obtained $f_{\rm{T,e}}$ values in Fig. \ref{fig:fT_fMAX_HF}, are comparable with $f_{\rm{T}}$ reported in \cite{WanWan14} at -\SI{1.8}{\volt}$\leq$$V_{\rm{GS}}$$\leq$-\SI{1.7}{\volt} and in \cite{LiTia18} at -\SI{1.2}{\volt}$\leq$$V_{\rm{GS}}$$\leq$-\SI{0.9}{\volt}. The bias-dependence observed for $f_{\rm{T,e}}$ and $f_{\rm{MAX,e}}$ is related to the $V_{\rm{G}}$-dependent $R_{\rm{C}}$ extracted for each device (cf. Fig. \ref{fig:RC_YFM_HF}).

Interestingly, the highest $f_{\rm{T,e}}$ has been obtained for the shortest device \cite{ZhuPar16} despite having the highest $R_{\rm{C}}$ among the three devices under study (see Fig. \ref{fig:RC_YFM_HF}). This result can be explained by an outstanding device electrostatics, i.e., the low values of $C_{\rm{gg,t}}$ and $C_{\rm{gd,t}}$ associated to the device reported in \cite{ZhuPar16} (see Table \ref{tab:Cg_param}) diminish the impact of $R_{\rm{C}}$ on $f_{\rm{T,e}}$. $R_{\rm{C}}$ extracted for the \SI{400}{\nano\meter}-long transistor \cite{WanWan14} is the lowest and most bias-independent at $V_{\rm{DS}}$=-\SI{0.1}{\volt} of the devices under study. Therefore, the device described in \cite{WanWan14} has the most linear $f_{\rm{T,e}}$ performance among the devices highlighted in Fig. \ref{fig:fT_fMAX_HF} (a), the non-optimal electrostatics of this device \cite{WanWan14}, i.e., large capacitance values (see Table \ref{tab:Cg_param}), hinders higher $f_{\rm{T,e}}$  values despite the lower $R_{\rm{C}}$ in comparison to \cite{ZhuPar16}. Fig. \ref{fig:fT_fMAX_HF} (b) shows that $f_{\rm{MAX,e}}$ in \cite{WanWan14} and \cite{LiTia18} are close to each other despite a lower $R_{\rm{C}}$ has been obtained for the \SI{300}{\nano\meter}-long device \cite{WanWan14} in comparison to the largest one \cite{LiTia18}. Furthermore, $R_{\rm{g}}$ is similar for both devices as shown in Table \ref{tab:Cg_param}. Hence, the large value of the capacitances obtained here for the device in \cite{WanWan14} compared to the ones obtained for \cite{LiTia18} impedes higher values of $f_{\rm{MAX,e}}$ for \cite{WanWan14}. It is important to highlight that a minimal change in these capacitances together with the $R_{\rm{C,YFM}}$ formed in the metal-channel interface, can modify the performance of these devices in HF. Therefore, in order to improve the dynamic HF performance of BPFETs, extreme care must be taken with $R_{\rm{C}}$ and the device electrostatics. The first parameter strongly influences not only the magnitude but the $f_{\rm{T,e}}$ and $f_{\rm{MAX,e}}$ response over bias, while the latter related parameters can define its value in combination with $R_{\rm{C}}$. $R_{\rm{g}}$ has not shown an important impact on the HF performance due to a lower sheet resistance of the metal gates as previously pointed out elsewhere \cite{YinAlM17}.

\section{Conclusion}
The drift-diffussion-based Y-function method has been used here to find the bias dependence of the $R_{\rm{C}}$ of different BPFET technologies, without the need of additional test structures or adjustment of a set of parameters, in contrast to other methods, i.e., TLM or CM. In general, extracted $R_{\rm{C}}$ values with Y-function are similar to the reference values, obtained with other costly and less straightforward extraction methods. Y-function method also captures the temperature dependence of $R_{\rm{C}}$ for a BP-device. Additionally, it has been found that the influence of the contact and channel resistance can be studied since YFM values are not affected by the latter one. Combined with other key parameters, namely, $g_{\rm{m}}$, $g_{\rm{ds}}$, $R_{\rm{g}}$ and intrinsic capacitances, it is possible to get the RF FoMs such as $f_{\rm{T,e}}$ and $f_{\rm{MAX,e}}$ which are usually strongly influenced by $R_{\rm{C}}$, especially at short channel lengths. YFM method is applicable to obtain $R_{\rm{C}}$ at different biases, potential contact barriers and temperature ranges, which means that it is an efficient and reliable methodology for data extraction based on the individual DC characteristics of BPFETs.


\ack
This project has been financially supported by the Instituto Polit\'ecnico Nacional, Mexico under the contract no. SIP/20200617 and from the European Union\textsc{\char13}s Horizon 2020 research and innovation programme under grant agreements No GrapheneCore2 785219 and No GrapheneCore3 881603, from Ministerio de Ciencia, Innovaci\'on y Universidades under grant agreement RTI2018-097876-B-C21(MCIU/AEI/FEDER, UE). This article has been partially funded by the European Regional Development Funds (ERDF) allocated to the Programa Operatiu FEDER de Catalunya 2014-2020, with the support of the Secretaria d’Universitats i Recerca of the Departament d’Empresa i Coneixement of the Generalitat de Catalunya for emerging technology clusters to carry out valorization and transfer of research results. Reference of the GraphCAT project: 001-P-001702.



\section*{References}
\begin{thebibliography}{10}
	
\bibitem{CaoW18} Cao W, Jiang J, Xie X, Pal A, Chu JH, Kang J, Banerjee K 2018 2-D Layered Materials for Next-Generation Electronics: Opportunities and Challenges, \emph{IEEE Transactions on Electron Devices}, \textbf{65}, 4109-4121.

\bibitem{FioGia15} Fiori G, Bonaccorso F, Iannaccone G, Palacios T, Neumaier D, Seabaugh A, Banerjee SK, Colombo L 2014 Electronics based on two-dimensional materials, \emph{Nature Nanotechnology}, \textbf{9}, 768-779.

\bibitem{LiYu14} Li L, Yu Y, Ye GJ, Ge Q, Ou X, Wu H, Feng D, Chen XH, Zhang Y 2014 Black Phosphorus Field-effect Transistors, \emph{Nature Nanotechnology}, \textbf{9}, 372-377.

\bibitem{LiYu19} Li X, Yu Z, Xiong X, Li T, Gao T, Wang R, Huang R, Wu Y 2019 High-speed black phosphorus field-effect transistors approaching ballistic limit, \emph{Science Advances}, \textbf{5}, aau3194.

\bibitem{ZhuPar16} Zhu W, Park S, Yogeesh MN, McNicholas KM, Bank SR, Akinwande D 2016 Black Phosphorus Flexible Thin Film Transistors at Gighertz Frequencies \emph{Nano Letters}, \textbf{16}, 2301-2306.

\bibitem{WanWan14} Wang H, Wang X, Xia F, Wang L, Jiang H, Xia Q, Chin ML, Dubey M, Han SJ 2014 Black Phosphorus Radio-Frequency Transistors, \emph{Nano Letters}, \textbf{14}, 6424-6429.

\bibitem{LiTia18} Li T, Tian M, Li S, Huang M, Xiong X, Hu Q, Li S, Li X, Wu Y 2018 Black Phosphorus Radio Frequency Electronics at Cryogenic Temperatures, \emph{Advanced Electronic Materials}, \textbf{4}, 1800138.

\bibitem{AshPen15} Penumatcha A, Salazar R, Appenzeller J 2015 Analysing black phosphorus transistors using an analytic Schottky barrier MOSFET model, \emph{Nature Communications}.

\bibitem{HarMat15} Haratipour N, Robbins MC, Koester SJ 2015 Black Phosphorus p-MOSFETs With 7-nm HfO2 Gate Dielectric and Low Contact Resistance, \emph{IEEE Electron Device Letters}, \textbf{36}, 411-413.

\bibitem{YanCha17} Yang L, Charnas A, Qiu G, Lin YM, Lu CC, Tsai W, Paduano Q, Snure M, Ye PD 2017 How Important Is the Metal-Semiconductor Contact for Schottky Barrier Transistors: A Case Study on Few-Layer Black Phosphorus?, \emph{ACS Omega}, \textbf{2}, 4173-4179.

\bibitem{YarHar19} Yarmoghaddam E, Haratipour N, Koester SJ, Rakheja S 2019 A virtual-source emission-diffusion I-V model for ultra-thin black phosphorus field-effect transistors, \emph{Journal of Applied Physics}, \textbf{125}, 165706.

\bibitem{YarHar20} Yarmoghaddam E, Haratipour N, Koester SJ, Rakheja S 2020 A Physics-Based Compact Model for Ultrathin Black Phosphorus FETs$-$Part I: Effect of Contacts, Temperature, Ambipolarity, and Traps, \emph{IEEE Transactions on Electron Devices}, \textbf{67}, 389-396.

\bibitem{YarHar19_II} Yarmoghaddam E, Haratipour N, Koester SJ, Rakheja S 2019 A Physics-Based Compact Model for Ultrathin Black Phosphorus FETs$-$Part II: Model Validation Against Numerical and Experimental Data, \emph{IEEE Transactions on Electron Devices}, \textbf{67}, 397-405.

\bibitem{Ghi88} Ghibaudo G 1988 New method for the extraction of MOSFET parameters, \emph{Electronics Letters}, \textbf{24}, 543-545.

\bibitem{ChaZhu14} Chang HY, Zhu W, Akinwande D 2014 On the mobility and contact resistance evaluation for transistors based on MoS2 or two-dimensional semiconducting atomic crystals, \emph{Applied Physics Letters}, \textbf{104}, 113504.

\bibitem{ParSon18} Park H, Son J, Kim J 2018 Reducing the contact and channel resistances of black phosphorus via low-temperature vacuum annealing, \emph{Journal of Materials Chemistry C}, \textbf{6}, 1567-1572.

\bibitem{PacFei20} Pacheco-Sanchez A, Feijoo PC, Jimenez D 2020 Contact resistance extraction of graphene FET technologies based on individual device characterization, \emph{submitted to Solid-State Electronics}.

\bibitem{PacCla16} Pacheco-Sanchez A, Claus M, Mothes S, Schr\"oter M 2016 Contact resistance extraction methods for short- and long-channel carbon nanotube field-effect transistors, \emph{Solid-State Electronics}, \textbf{125}, 161-166.

\bibitem{HaoCab85} Hao C, Cabon-Till B, Cristoloveanu S, Ghibaudo G, Experimental determination of short-channel MOSFET parameters, \emph{Solid-State Electronics}, \textbf{28}, 1025-1030.

\bibitem{PacCla20} Pacheco-Sanchez A, Claus M 2020 Bias-Dependent Contact Resistance Characterization of Carbon Nanotube FETs, \emph{IEEE Transactions on Nanotechnology}, \textbf{19}, 47-51.

\bibitem{HarNam16} Haratipour N, Namgung S, Oh SH, Koester SJ 2016 Fundamental Limits on the Subthreshold Slope in Schottky Source/Drain Black Phosphorus Field-Effect Transistors, \emph{ACS Nano}, \textbf{10}, 3791-3800.

\bibitem{YinAlM17} Yin D, AlMutairi A, Yoon Y 2017 Assessment of High-Frequency Performance Limit of Black Phosphorus Field-Effect Transistors, \emph{IEEE Transactions on Electron Devices}, \textbf{64}, 2984-2991.

\bibitem{PacJim20} Pacheco-Sanchez A and Jim\'enez D, On the accuracy of Y-function methods for parameters extraction of two-dimensional FETs across different technologies, \emph{submitted to Electronics Letters}.

\bibitem{XiaYin15} Chen X, Wu Y, Wu Z, Han Y, Xu S, Wang L, Ye W, Han T, He Y, Cai Y, Wang N 2015 High-quality sandwiched black phosphorus heterostructure and its quantum oscillations, \emph{Nat. Commun.}, \textbf{6}, 7315.

\bibitem{CheKua20} Li C, Xiong K, Li L, Guo Q, Chen X, Madjar A, Watanabe K, Taniguchi T, Hwang J. C. M, Xia F 2020 Black Phosphorus High-Frequency Transistors with Local Contact Bias, \emph{ACS Nano}, \textbf{14}, 2118–2125.

\end{thebibliography}
\end{document}